\documentclass[12pt,a4paper]{article}
\usepackage{amssymb}
\usepackage{amsmath}
\RequirePackage{graphicx}
\RequirePackage{mathptmx}      
\RequirePackage{flushend}
\RequirePackage[numbers,sort&compress]{natbib}
\usepackage[colorlinks]{hyperref}
\hypersetup{
     breaklinks=true,
     pdfstartview={FitH},    
    colorlinks=true,       
    linkcolor=blue,          
    citecolor=red,        
    filecolor=magenta,      
    urlcolor=Green,           
    anchorcolor=blue,      
    linktocpage=true
}
\usepackage{psfig}
\begin{document}
\begin{center}{\Large\bf Constraints on reconstructed dark energy model from SN Ia and BAO/CMB observations}
\\[15mm]
Abdulla Al Mamon,$^{a,}$\footnote{Present affiliation:\\ Manipal Centre for Natural Sciences, Manipal University,\\ Manipal-576104, Karnataka, India.}$^{,}$\footnote{E-mail : abdullaalmamon.rs@visva-bharati.ac.in, abdulla.mamon@manipal.edu}~
Kazuharu Bamba$^{b,}$\footnote{E-mail:  bamba@sss.fukushima-u.ac.jp}~and Sudipta Das$^{a,}$\footnote{E-mail:  sudipta.das@visva-bharati.ac.in}\\
{\em ${}^{a}$Department of Physics, Visva-Bharati, Santiniketan- 731235, ~India.}\\
{\em ${}^{b}$Division of Human Support System, Faculty of Symbiotic Systems Science,\\ Fukushima University, Fukushima 960-1296, ~Japan}\\
[15mm]
\end{center}
\vspace{0.5cm}
\vspace{0.5cm}
\pagestyle{myheadings}
\newcommand{\be}{\begin{equation}}
\newcommand{\ee}{\end{equation}}
\newcommand{\bea}{\begin{eqnarray}}
\newcommand{\eea}{\end{eqnarray}}
\newcommand{\bc}{\begin{center}}
\newcommand{\ec}{\end{center}}
\begin{abstract}
The motivation of the present work is to reconstruct a dark energy model through the {\it dimensionless dark energy function} $X(z)$, which is the dark energy density in units of its present value. In this paper, we have shown that a scalar field $\phi$ having a phenomenologically chosen $X(z)$ can give rise to a transition from a decelerated to an accelerated phase of expansion for the universe. We have examined the possibility of constraining various cosmological parameters (such as the deceleration parameter and the effective equation of state parameter) by comparing our theoretical model with the latest Type Ia Supernova (SN Ia), Baryon Acoustic Oscillations (BAO) and Cosmic Microwave Background (CMB) radiation observations. Using the joint analysis of the SN Ia+BAO/CMB dataset, we have also reconstructed the scalar potential from the parametrized $X(z)$. The relevant potential is found, which comes to be a polynomial in $\phi$. From our analysis, it has been found that the present model favors the standard $\Lambda$CDM model within $1\sigma$ confidence level.
\end{abstract}
\section{Introduction}
The various cosmological observations such as Type Ia Supernovae~\cite{SN,Riess:1998cb}, cosmic microwave background (CMB) radiation~\cite{Planck:2015xua, Ade:2015lrj, Ade:2014xna, Ade:2015tva, Array:2015xqh, Komatsu:2010fb, Hinshaw:2012aka}, 
large scale structure~\cite{LSS,Seljak:2004xh}, baryon acoustic oscillations (BAO)~\cite{Eisenstein:2005su}, and weak lensing~\cite{Jain:2003tba} 
have supported that the expansion of the current universe is accelerating. All of these observations also strongly indicate that the alleged acceleration is rather a recent phenomenon and the universe was decelerating in the past. Two representative approaches have been proposed to account for the late-time cosmic acceleration. The first approach is to assume the existence of ``dark energy'' (DE) in the framework of general relativity. The second approach is to consider the modification of gravity on the large scale (for reviews on the issues of DE and the modified theories of gravitation, see, for example, \cite{Nojiri:2010wj,Nojiri:2006ri,Book-Capozziello-Faraoni,Capozziello:2011et,delaCruzDombriz:2012xy,Bamba:2012cp,Joyce:2014kja,
Koyama:2015vza,Bamba:2015uma}). In this work, we will concentrate only on the first approach and consider DE to be responsible for this accelerated phenomenon. There are some excellent review articles where various DE models have been comprehensively discussed \cite{3,4,4a,5}. The simplest candidate of DE a is cosmological constant $\Lambda$ whose energy density remains constant with time and its {\it equation of state} (EoS) parameter is, $\omega_{\Lambda}=-1$. However, the models based upon cosmological
constant suffer from the {\it fine tuning} and the cosmological {\it coincidence} problems \cite{sw, Steinhardt}. Scalar field models with generic features can alleviate these problems and provide the late-time evolution of the universe (see Ref. \cite{4} for a review). Scalar field models are very popular as the simplest generalization of cosmological constant is provided by a scalar field, dubbed as quintessence field, which can drive the acceleration with some suitably chosen potentials. In this case, one needs some degree of fine tuning of the initial conditions to account for the accelerated expansion of the universe and none of the potentials really have proper theoretical support from field theory explaining their origin (for review, see \cite{6}). In the last decade, an enormous number of DE models were explored to explain the origin of this late time acceleration of the universe and none of these models have very strong observational evidence \cite{4}. Therefore, the search is on for a suitable DE model and the present study is one of them.\\
\par In Ref. \cite{ellis}, Ellis and Madsen had discussed about {\it reconstruction} method to find the scalar field potential. Recently, this method finds a very wide application in current research in cosmology. However, there are two types of reconstruction, namely, parametric and non-parametric. The parametric reconstruction method is an attempt to build up a model by assuming a specific evolution scenario for a model parameter and then estimate the values of the parameters from different observational datasets. On the other hand, the non-parametric reconstruction method does not require any specific assumption for the parameters and finds the nature of
cosmic evolution directly from observational dataset.\\
\par In the context of DE, the reconstruction method was first discussed in \cite{starpr}, where Starobinsky determined the scalar field potential from the observational dataset from the behavior of density perturbations in dust-like matter. Some other earlier works on reconstruction have been discussed in \cite{tdsprl,relz} where the dataset of cosmological distance measurement has been used. In practice, a large number of dynamical models have been proposed for DE in which the properties of DE component are generally summarized as a perfect fluid with a time-dependent EoS parameter $\omega_{\phi}(z)$. In building up the DE model by the parametric reconstruction method, efforts are normally made through the DE EoS parameter. In literature, there are many examples where the authors had proposed different redshift parametrizations of $\omega_{\phi}$ to fit with observational data \cite{lz1,lz2,cpl1,cpl2} (for review, see also Refs. \cite{jmang,kshi,aam,aam2}). However, it has been found that the parametrization of the energy density $\rho_{\phi}(z)$ (which depends on its EoS parameter through an integral) provides tighter constraints than $\omega_{\phi}(z)$ from the same observational dataset (for details, see Refs.\cite{mt0,mt1,mt2}). Recently, many investigations have been performed to find the actual functional form of $\omega_{\phi}$ directly from the available datasets \cite{np1,np2,np3,np4}. However, the problem with this method is that the parameters of interest usually contain noisy data. The present work uses the idea of parametrizing the DE density, where we have presented a parametric reconstruction of the DE function $X(z)$ (which is basically the DE density in units of its present value) to study the essential properties of DE. The basic properties of this chosen $X(z)$ has been discussed in detail in the next section. The functional form of $X(z)$ depends on the model parameters which have been constrained from the observational datasets. The constraints on the model parameters are obtained by using various observational datasets (namely, SN Ia, BAO and CMB) and $\chi^{2}$ minimization technique. With the estimated values of model parameters, we have then reconstructed the deceleration parameter and the EoS parameter at the $1\sigma$ and $2\sigma$ confidence levels. Furthermore, we have also tried to reconstruct the scalar potential $V(\phi)$ directly from the dark energy function $X(z)$. Clearly, the present study enables us to construct the scalar field potential without assuming its functional form. This is one of the main objectives of the present work. We have found that the results obtained in this work are consistent with the recent observations and the model do not deviate very far from the $\Lambda$CDM model at the present epoch. \\
\par The outline of the paper is as follows. In the next section, we have presented the basic formalism of a flat FRW cosmology along with the definitions of different cosmological parameters. We have then solved the field equations for this toy model using a specific choice of the dark energy function $X(z)$. The observational datasets and methodology are discussed in section \ref{data}. The main results of this analysis are summarized in section \ref{result}. Finally, in the last section, we have presented our main conclusions.
\section{Field equations and their solutions}\label{sec2}
The action for a scalar field $\phi$ and the Einstein-Hilbert term is 
described as 
\begin{equation}
S = \int d^4 x \sqrt{-g} \left( \frac{R}{2\kappa^2} 
 - \frac{1}{2}\partial_\mu \phi \partial^\mu \phi - V(\phi) \right)\,, 
\label{eq:2.1}
\end{equation}
where $g$ is the determinant of the metric $g_{\mu\nu}$ and  $R$ is the scalar curvature.  
In this work, we have chosen natural units in which $\kappa^2=8\pi G = 1$. We have assumed the spatially flat Friedmann-Lema\^{i}tre-Robertson-Walker (FLRW) 
space-time 
\begin{equation}
ds^2 = dt^2 - a^2(t) \sum_{i=1,2,3}\left(dx^i\right)^2 \,. 
\label{eq:2.2}
\end{equation}
Here, $a$ is the scale factor of the universe (taken to be $a = 1$ at the present epoch). In the above background, the corresponding Einstein field equations can be obtained as,
\be\label{fe1}
3H^{2}=\rho_{m} + \frac{1}{2}{\dot{\phi}}^{2} + V(\phi) 
\ee
\be\label{fe2}
2{\dot{H}} + 3H^{2}=-\frac{1}{2}{\dot{\phi}}^{2} + V(\phi) 
\ee
where $H=\frac{\dot{a}}{a}$ is the Hubble parameter, $\rho_{m}$ is the energy density of the matter field and $\phi$ is the scalar field with potential $V(\phi)$. Here and throughout the paper, an overhead dot implies differentiation with respect to the cosmic time $t$. \\
From equations (\ref{fe1}) and (\ref{fe2}), one can note that the energy density $\rho_{\phi}$ and pressure $p_{\phi}$ of the scalar field $\phi$ are given by
\be\label{eqrp1}
\rho_{\phi} = \frac{1}{2}{\dot{\phi}}^{2} + V(\phi) 
\ee
\be\label{eqpp1}
p_{\phi}=\frac{1}{2}{\dot{\phi}}^{2} - V(\phi) 
\ee
Also, the conservation equation for the scalar field $\phi$ takes the form
\be\label{ce1}
{\dot{\rho}}_{\phi} + 3H(\rho_{\phi} + p_{\phi})=0  
\ee
From these equations, one can now easily arrive at the matter conservation equation as
\be\label{ce2}
{\dot{\rho}}_{m} + 3H\rho_{m}=0 
\ee
which can be easily integrated to yield
\be\label{cem2}
\rho_{m}=\rho_{m0}a^{-3} 
\ee
where $\rho_{m0 }$is an integrating constant which denotes the present value of the matter energy density. 
From equation (\ref{ce1}), the corresponding EoS parameter can be written as
\be\label{wphirho} 
\omega_{\phi}(z)=\frac{p_{\phi}}{\rho_{\phi}}=-1-\frac{(1+z)}{3X(z)}\frac{dX(z)}{dz}
\ee
so that,
\be
X(z)={\rm exp}{\left[3\int^{z}_{0}(1+\omega_{\phi}(z^{\prime}))dln(1+z^{\prime})\right]} 
\ee
where, $X(z)=\frac{\rho_{\phi}(z)}{\rho_{\phi 0}}$, $\rho_{\phi 0}$ denotes the present value of $\rho_{\phi}(z)$ and $z$ is the redshift parameter which is given by
$z=\frac{1}{a} -1$. It is evident from equation (\ref{wphirho}) that the EoS parameter becomes cosmological constant ($\omega_{\phi}=-1$) when $X(z)=$ constant. Clearly, the quantity $X(z)$, instead of $\omega_{\phi}(z)$, is a very good probe to investigate the nature of dark energy. In Refs. \cite{mt0,mt2}, the authors argued that one can obtain more information by reconstructing $\rho_{\phi}(z)$ rather than $\omega_{\phi}(z)$ from the observational data.\\
\par Using equations (\ref{fe1}), (\ref{cem2}) and (\ref{eans1}), the Hubble parameter for this model can be written as
\be\label{eh}
H(z)=H_{0}\sqrt{\left[ \Omega_{m0}(1+z)^{3} + \Omega_{\phi 0}X(z)\right]}
\ee
where $H_{0}$ is the present value of $H(z)$, $\Omega_{m0}=\frac{\rho_{m0}}{3H^{2}_{0}}$ and $\Omega_{\phi 0}=\frac{\rho_{\phi 0}}{3H^{2}_{0}}=(1-\Omega_{m0})$ are the present value of the density parameters of matter and scalar field respectively.\\ 
Next, we have used this $H$ to find out the behavior of the deceleration parameter $q$, which is defined as
\be\label{eq}
q=-\frac{\ddot{a}}{aH^{2}}=-{\left(1+\frac{\dot{H}}{H^{2}}\right)}
\ee
where $\dot{H}=-(1+z)H\frac{dH}{dz}$. \\
Using equations (\ref{eh}) and (\ref{eq}), we have obtained the expressions for the deceleration parameter $q$ (in terms of redshift $z$) as,
\be\label{eq2}
q(z)=-1 + \frac{(1+z){\left[3\Omega_{m0}(1+z)^{2}+(1-\Omega_{m0})\frac{dX(z)}{dz}\right]}}{2{\left[\Omega_{m0}(1+z)^{3}+(1-\Omega_{m0})X(z)\right]}}
\ee
Combining equations (\ref{eqrp1}) and (\ref{eqpp1}), one can obtain an expression for the scalar field $\phi(z)$ as 
\bea\label{eqrcpz1}
{\left(\frac{d\phi}{dz}\right)}^{2}=\frac{3H^{2}_{0}(1-\Omega_{m0})}{(1+z)H^{2}(z)}\frac{dX(z)}{dz}\nonumber\\ \\
\Rightarrow~\phi(z)=\phi_{0}+\int_{z}{\sqrt{\frac{3H^{2}_{0}(1-\Omega_{m0})}{(1+z^{\prime})H^{2}(z^{\prime})}\frac{dX(z^{\prime})}{dz^{\prime}}}}dz^{\prime}
\eea
where $\phi_{0}$ is an integration constant.\\
Similarly, using equations (\ref{eqrp1}) and (\ref{eqpp1}), one can reconstruct the potential for the scalar field as
\be\label{eqrcvz1}
V(z)=3H^{2}_{0}(1-\Omega_{m0}){\left[X(z)-\frac{(1+z)}{6}\frac{dX(z)}{dz}\right]}
\ee
Therefore, we can obtain the expression for the potential $V(\phi)$ as a function of $\phi$, by solving equations (\ref{eqrcpz1}) and (\ref{eqrcvz1}) if the values of the model parameters and the functional form of $X(z)$ are given.\\
\par Now, out of four equations (\ref{fe1}), (\ref{fe2}), (\ref{ce1}) and (\ref{ce2}), only three are independent as any one of them can be derived from the Einstein field equations with the help of the other three in view of the Bianchi identities. So, we have four unknown parameters (namely, $H$, $\rho_{m}$, $\phi$ and $V(\phi)$) to solve for. Hence, in order to solve the system completely, we need an additional input. For the present work, we have considered a simple assumption regarding the functional form for the evolution of $X(z)$ and is given by
\be\label{eans1}
X(z)=(1+z)^{\alpha}{\rm e}^{\beta z}
\ee
where $\alpha$ and $\beta$ are arbitrary constants to be fixed by observations. For this choice of $X(z)$, the EoS parameter $\omega_{\phi}(z)$ comes out as
\be\label{wphiabl}
\omega_{\phi}(z)=-1+\frac{\alpha}{3}+\frac{\beta}{3}(1+z)
\ee
which is similar to the well-known linear redshift parametrization of the EoS parameter $\omega_{\phi}(z)$ given by \cite{lz1,lz2} 
\be 
\omega_{\phi}(z)=\omega_{0}+\omega_{1}z
\ee
for $\omega_{0}=\omega_{\phi}(z=0)=-1+\frac{\alpha}{3}+\frac{\beta}{3}$ and $\omega_{1}=\frac{\beta}{3}$. This parametrization is well behaved at low redshifts, but 
it diverges at high redshift. However, the above choice of $\omega_{\phi}(z)$ has been widely used in the context of dark energy (as it is a late-time phenomenon),
due to its simplicity. When $\alpha=0$ and $\beta=0$, the EoS parameter (\ref{wphiabl}) reduces to the standard $\Lambda$CDM model as well. Therefore, the simplicity of the functional form of $X(z)$ (or, equivalently, $\omega_{\phi}(z)$) makes it very attractive to study. In other words, the choice (\ref{eans1}) can be thought of as the parametrization of the DE density instead of $\omega_{\phi}(z)$. If desired cosmological scenario is achieved with this choice of $X(z)$, then some clues about the nature of DE may be obtained.\\
For this specific choice, equation (\ref{eh}) can be written as
\be
H(z)=H_{0}\sqrt{\left[ \Omega_{m0}(1+z)^{3} + \Omega_{\phi 0}(1+z)^{\alpha}{\rm e}^{\beta z}\right]}
\ee
The effective EoS parameter can be expressed in terms of $H$ and its derivative with respect to $z$ as,
\bea
\omega_{eff}(z)=\frac{p_{\phi}}{\rho_{m}+\rho_{\phi}}~~~~~~~~~~~~~~~~~~~~ \nonumber \\ \\
\Rightarrow ~\omega_{eff}(z)=-\frac{2{\dot{H}}+3H^{2}}{3H^{2}}=-1+\frac{2(1+z)}{3H(z)}\frac{dH(z)}{dz} 
\eea
and for the present model, the expression is
\be
\omega_{eff}(z)=-\frac{(1-\Omega_{m0})(3-\alpha -\beta -\beta z){\rm e}^{\beta z}(1+z)^{\alpha}}{3{\rm e}^{\beta z}(1+z)^{\alpha}-3\Omega_{m0}((1+z)^{3}-{\rm e}^{\beta z}(1+z)^{\alpha})}
\ee \\
In this case, $q(z)$, $V(z)$ and $\phi(z)$ evolve as
\be\begin{split} &
q(z)=-1 \\& + \frac{{\left[3\Omega_{m0}(1+z)^{3}+(1-\Omega_{m0}){\rm e}^{\beta z}\lbrace \alpha (1+z)^{\alpha}+\beta (1+z)^{\alpha +1}\rbrace\right]}}{2{\left[\Omega_{m0}(1+z)^{3}+(1-\Omega_{m0})(1+z)^{\alpha}{\rm e}^{\beta z}\right]}}
\end{split}
\ee \\
\be\label{pz1}
\phi(z)=\phi_{0} + \int_{z} \sqrt{\frac{{(1-\Omega_{m0}){\rm e}^{\beta z^{\prime}}(1+z^{\prime})^{\alpha}[\alpha +\beta(1+z^{\prime})]}}{{\Omega_{m0}(1+z^{\prime})^{3}+(1-\Omega_{m0}){\rm e}^{\beta z^{\prime}}(1+z^{\prime})^{\alpha}}}}\frac{dz^{\prime}}{(1+z^{\prime})}
\ee \\
\be\label{vz1}
V(z)=\frac{H^{2}_{0}(\Omega_{m0}-1)}{2}{\rm e}^{\beta z}(1+z)^{\alpha}[\alpha +\beta(1+z)-6]
\ee \\
Before reconstructing the functional form for $V(\phi)$ for given values of the model parameters (e.g., $\alpha$ and $\beta$), we first obtain the allowed ranges for these parameters from the observational datasets. In the next section, we shall attempt to estimate the values of $\alpha$ and $\beta$ using available observational datasets, so that the said model can explain the evolution history of the universe more precisely.
\section{Data analysis methods}\label{data}
Here, we have explained the method employed to constrain the theoretical models by using the recent observational datasets from Type Ia Supernova (SN Ia), Baryon Acoustic Oscillations (BAO) and Cosmic Microwave Background (CMB) radiation data surveying. We have used the $\chi^{2}$ minimum test with these datasets and found the best fit values of
arbitrary parameters for $1\sigma$ and $2\sigma$ confidence levels (as discussed in section \ref{result}). In the following subsections, the $\chi^{2}$ analysis used for those datasets is described.
\subsection{SN Ia}
Firstly, we have used recently released Union2.1 compilation data \cite{sn1agnc} of 580 data points which has been widely used in recent times to constraint different dark energy models. The $\chi^2$ function for the SN Ia dataset is given by \cite{sndatamethodgnc}
\be\label{eqchisnia} 
\chi^2_{SN Ia}= {P} - \frac{{Q}^2}{R}
\ee
where $P$, $Q$ and $R$ are defined as follows
\bea
P = \sum^{580}_{i=1} \frac{[{\mu}^{obs}(z_{i}) - {\mu}^{th}(z_{i})]^2}{\sigma^2_{\mu}(z_{i})}\\
Q= \sum^{580}_{i=1} \frac{[{\mu}^{obs}(z_i) - {\mu}^{th}(z_{i})]}{\sigma^2_{\mu}(z_{i})}
\eea
and
\be 
R= \sum^{580}_{i=1} \frac{1}{\sigma^2_{\mu}(z_{i})}
\ee 
where $\mu^{obs}$ represents the observed distance modulus while $\mu^{th}=5{\rm log}_{10}{[(1+z)\int^{z}_{0}\frac{H_{0}}{H(z)}dz]}+25-5{\rm log}_{10}H_{0}$, is the corresponding theoretical one. Also, the quantity $\sigma_{\mu}$ represents the statistical uncertainty in the distance modulus.\\ 
\par Alternatively, $\chi^2_{SN Ia}$ can be written (in terms of covariance matrix) as
\be
\chi^2_{SN Ia}=X^{T} {C}^{-1}X\nonumber
\ee
where $X$ is a vector of differences $X_{i}=\mu^{th}(z_{i})-\mu^{obs}(z_{i})$, and ${C}^{-1}$ is the inverse Union 2.1 compilation covariance matrix. It deserves mention that for large sample sets, one can use either equation (\ref{eqchisnia}) or equation (\ref{eqchibaocmb}) without any loss of generality.
\subsection{BAO/CMB}
Next, we have considered BAO \cite{bao1,bao2,bao3} and CMB \cite{cmb} measurement dataset to obtain the BAO/CMB constraints on the model parameters. In Ref. \cite{baocmb2}, the authors have obtained the BAO/CMB constrains on the model parameters by considering only two BAO measurements, whereas here we have considered six BAO data points (see table \ref{baodata}). For BAO dataset, the results from the WiggleZ Survey \cite{bao3}, SDSS DR7 Galaxy sample \cite{bao2} and 6dF Galaxy Survey \cite{bao1} datasets have been used. On the other hand, the CMB measurement considered is derived from the WMAP7 observations \cite{cmb}. The discussion about the BAO/CMB dataset has also been presented in a very similar way in \cite{aam3}, but the details of methodology for obtaining the BAO/CMB constraints on model parameters is available in Ref. \cite{goistri}. For this dataset, the $\chi^2$ function is defined as \cite{goistri}
\be\label{eqchibaocmb}
\chi^2_{BAO/CMB} = X^{T}C^{-1}X
\ee
where
\be
X=\left( \begin{array}{c}
        \frac{d_A(z_\star)}{D_V(0.106)} - 30.95 \\
        \frac{d_A(z_\star)}{D_V(0.2)} - 17.55 \\
        \frac{d_A(z_\star)}{D_V(0.35)} - 10.11 \\
        \frac{d_A(z_\star)}{D_V(0.44)} - 8.44 \\
        \frac{d_A(z_\star)}{D_V(0.6)} - 6.69 \\
        \frac{d_A(z_\star)}{D_V(0.73)} - 5.45
        \end{array} \right)\,,
\ee
and
\bc
$C^{-1}={\left(
\begin{array}{cccccc}
 0.48435 & -0.101383 & -0.164945 & -0.0305703 & -0.097874 & -0.106738 \\
 -0.101383 & 3.2882 & -2.45497 & -0.0787898 & -0.252254 & -0.2751 \\
 -0.164945 & -2.45499 & 9.55916 & -0.128187 & -0.410404 & -0.447574 \\
 -0.0305703 & -0.0787898 & -0.128187 & 2.78728 & -2.75632 & 1.16437 \\
 -0.097874 & -0.252254 & -0.410404 & -2.75632 & 14.9245 & -7.32441 \\
 -0.106738 & -0.2751 & -0.447574 & 1.16437 & -7.32441 & 14.5022
\end{array}
\right)}$
\ec
\begin{table*}
\begin{center}
\caption{\it Values of $\frac{d_A(z_\star)}{D_V(Z_{BAO})}$ for different values of $z_{BAO}$. Here, $d_A(z)=\int_0^z \frac{dz'}{H(z')}$ is the co-moving angular-diameter
distance, $z_\star \approx 1091$ is the decoupling time and $D_V(z)=\left[d_A(z)^2\frac{z}{H(z)}\right]^{\frac{1}{3}}$ is the dilation scale \cite{goistri}.}
\begin{tabular}{|c||c|c|c|c|c|c|}
\hline
$z_{BAO}$  & 0.106  & 0.2& 0.35 & 0.44& 0.6& 0.73\\
\hline
$\frac{d_A(z_\star)}{D_V(Z_{BAO})}$ &  $30.95 \pm 1.46$& $17.55 \pm 0.60$
& $10.11 \pm 0.37$ & $8.44 \pm 0.67$ & $6.69 \pm 0.33$ & $5.45 \pm 0.31$
\\
\hline
\end{tabular}
\label{baodata}
\end{center}
\end{table*}
\par In this work, we have also considered the CMB shift parameter data (which is derived from Planck observation \cite{planckR}) and have examined its impact on the present dark energy constraints. For this dataset, the details of the methodology for obtaining the constraints on model parameters are described in Ref. \cite{planckR}). \\
\par Hence, the total $\chi^2$ for the combined dataset (SNIa+BAO/CMB) is given by 
\be 
\chi^2_{tot}= \chi^2_{SN Ia} + \chi^2_{BAO/CMB}
\ee
For the combination of SN Ia and BAO/CMB datasets, one can now obtain the best-fit values of parameters by minimizing $\chi^2_{tot}$. Then, one can use the maximum likelihood method and take the total likelihood function ${\cal L}_{tot}={\rm e}^{-\frac{\chi^2_{tot}}{2}}$ as the products of these individual likelihood functions of each dataset, i.e., ${\cal L}_{tot}={\cal L}_{SN} \times {\cal L}_{BAO/CMB}$. The best-fit parameter values $b^{*}$ are those that maximize the likelihood function ${\cal L}_{tot}(b^{*})$, or equivalently minimize $\chi^2_{tot}(b^{*})=-2{\rm ln}{\cal L}_{tot}(b^{*})$. The contours of $1\sigma$ and $2\sigma$ constraints correspond to the sets of cosmological parameters (centered on $b^{*}$) bounded by $\chi^2_{tot}(b) = \chi^2_{tot}(b^{*}) + 2.3$ and $\chi^2_{tot}(b)= \chi^2_{tot}(b^{*}) + 6.17$ respectively. For the present model, we have minimized the $\chi^{2}$ function with respect to the model parameters $\lbrace \alpha ,\beta\rbrace$ to obtain their best fit values. In order to do so, we have fixed $\Omega_{m0}$ to some constant value.
\section{Results of data analysis}\label{result} 
Following the $\chi^{2}$ analysis (as presented in section \ref{data}), in this section, we have obtained the constraints on the model parameters $\alpha$ and $\beta$ for the combined dataset (SN Ia+BAO/CMB). In this work, we have obtained the confidence region ellipses in the $\alpha - \beta$ parameter space by fixing $\Omega_{m0}$ to $0.26$, $0.27$ and $0.28$ for the combined dataset. The $1\sigma$ and $2\sigma$ confidence level contours in $\alpha - \beta$ plane is shown in figure {\ref{figc} for SN Ia+BAO/CMB dataset. It has also been found from figure {\ref{figc} that current constraints favor a $\Lambda$CDM model within $1\sigma$ confidence limit (as shown by red dot). The best-fit values of $\alpha$ and $\beta$ are presented in the table \ref{tab:fntable1}. 
\begin{figure}[ht]
\begin{center}
\includegraphics[width=0.38\textwidth,height=0.21\textheight]{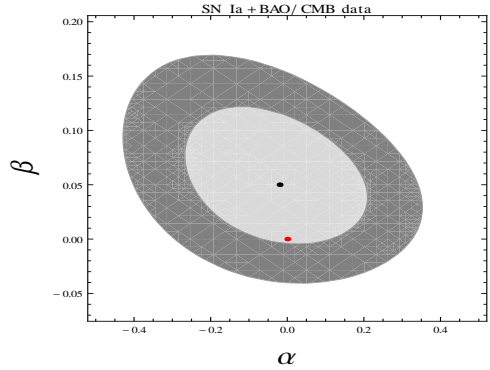}
\caption{\em Plot of $1\sigma$ (light gray) and $2\sigma$ (gray) confidence contours on $\alpha-\beta$ parameter space for SN Ia+BAO/CMB dataset. In this plot, black dot represents the best-fit value of the pair ($\alpha,\beta$) arising from the analysis of SN Ia+BAO/CMB dataset. Also, the red dot represents the standard $\Lambda$CDM model (as $\omega_{\phi}=-1$ for $\alpha=0$ and $\beta=0$). The plot is for $\Omega_{m0} = 0.27$. }
\label{figc}
\end{center}
\end{figure}
\begin{table}[ht]
\caption{\em Best fit values of the model parameters $\alpha$ and $\beta$ (within $1\sigma$ confidence level) for the analysis of SN Ia+BAO/CMB dataset with different choices of $\Omega_{m0}$. Here, $\chi^{2}_{m}$ represents the minimum value of $\chi^{2}$. }
\begin{center}
\begin{tabular}{|c|c|c|c|c|}
\hline
Data & $\Omega_{m0}$ & $\alpha$ & $\beta$ & $\chi^{2}_{m}$ \\ \hline
SN Ia+BAO/CMB & $0.26$ &$0.03$ & $0.07$ & $564.81$\\ 
& $0.27$ &$-0.02$ & $0.05$ & $564.79$\\
& $0.28$ &$-0.08$ & $0.03$ & $564.80$\\ 
\hline
\end{tabular}
\label{tab:fntable1} 
\end{center} 
\end{table}
The Marginalized likelihoods for the present model is shown in figure \ref{figl}. It is evident from the likelihood plots that the likelihood functions are well fitted
to a Gaussian distribution function for SN Ia+BAO/CMB dataset. For a comprehensive analysis, we have also used $\Omega_{m0}$ and $H_{0}$ as free parameters along with $\alpha$ and $\beta$. The result of corresponding statistical analysis is presented in table \ref{tab:fntable2}. It is clear from table \ref{tab:fntable1} and \ref{tab:fntable2} that the best-fit value of $\Omega_{m0}$ comes out to be $0.28$ which was one of the choices in table \ref{tab:fntable1}, and the corresponding values of $\alpha$ and $\beta$ does not differ by very large values. However, the values of the parameters ($H_{0}$ and $\Omega_{m0}$) obtained in the present work are slightly lower than the values obtained by the Planck analysis, which puts the limit on the parameters as, $H_{0}=67.3\pm 1.2$ km s$^{-1}$ Mpc$^{-1}$ and $\Omega_{m0}=0.315\pm 0.017$ with $1\sigma$ errors \cite{planckh0}.\\
\begin{figure}[ht]
\begin{center}
\includegraphics[width=0.38\textwidth,height=0.21\textheight]{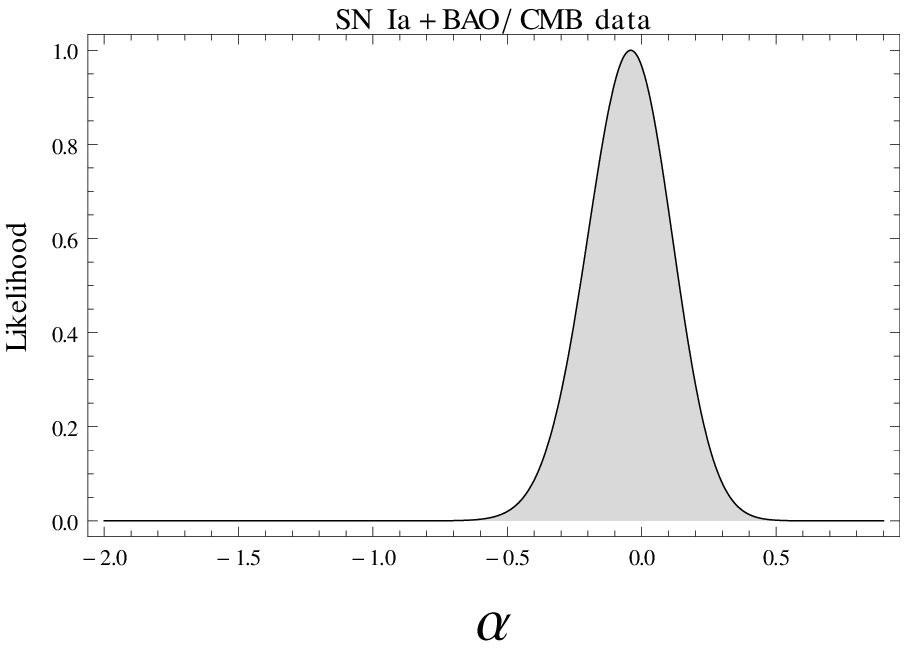}\\
\vspace{7mm}
\includegraphics[width=0.38\textwidth,height=0.21\textheight]{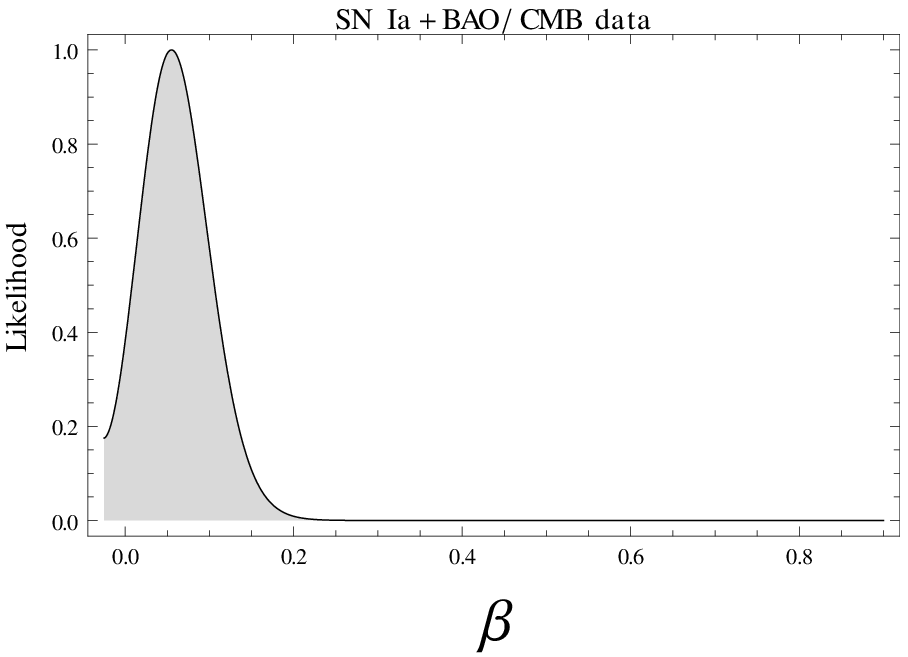}
\caption{\em The marginalised likelihood functions of the present model are shown for SN Ia+BAO/CMB dataset. Both the plots are for $\Omega_{m0} = 0.27$. }
\label{figl}
\end{center}
\end{figure}
\par In addition to this, we have also obtained the constraints on model parameters using the combination of SN Ia, BAO and the CMB shift parameter (which is derived from Planck observation \cite{planckR}) datasets to study the properties of our model extensively. For the SN Ia+BAO+CMB(Planck) dataset, the $1\sigma$ and $2\sigma$ confidence level contours in $\alpha - \beta$ plane is shown in figure {\ref{figcp}. The results of corresponding data analysis are summarized in table \ref{tab:fntable3} and \ref{tab:fntable4}. It has been found from figure \ref{figc} and \ref{figcp} that the constraints obtained on the parameter values by the SN Ia+BAO+CMB(Planck) dataset are very tight as compared to the constraints obtained from the SN Ia+BAO/CMB(WMAP7) dataset. However, the change in the best fit values of the model parameters ($\alpha$ and $\beta$) for the two datasets is very small. Also, the best fit values of $H_{0}$ and $\Omega_{m0}$ obtained in this case are very close to the values obtained by the Planck analysis \cite{planckh0}. We have found from figure \ref{figcp} that the present constraints obtained from SNIa+BAO+CMB(Planck) dataset favor a standard $\Lambda$CDM model within $2\sigma$ confidence limit as shown by the red dot, whereas for SN Ia+BAO/CMB dataset, the $\Lambda$CDM model was favored with $1\sigma$ confidence limit as evident from figure \ref{figc}.\\
\begin{figure}[ht]
\begin{center}
\includegraphics[width=0.38\textwidth,height=0.21\textheight]{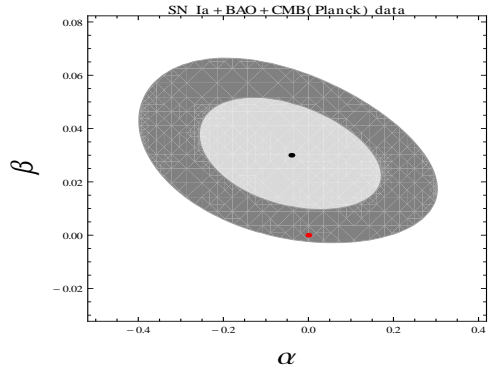}
\caption{\em This figure shows the $1\sigma$ (light gray) and $2\sigma$ (gray) confidence contours on $\alpha-\beta$ parameter space using the SN Ia+BAO+CMB(Planck) dataset. In this plot, the black dot represents the best-fit value of the pair ($\alpha,\beta$), whereas the red dot represents the standard $\Lambda$CDM model ($\alpha=0$ and $\beta=0$). The plot is for $\Omega_{m0} = 0.3$. }
\label{figcp}
\end{center}
\end{figure}
\begin{table}[ht]
\caption{\em Best fit values of $H_{0}$, $\Omega_{m0}$, $\alpha$ and $\beta$ (within $1\sigma$ confidence level) for the analysis of SN Ia+BAO/CMB dataset.}
\begin{center}
\begin{tabular}{|c|c|c|c|c|c|}
\hline
Data & $H_{0}$ &$\Omega_{m0}$ & $\alpha$ & $\beta$ & $\chi^{2}_{m}$ \\ \hline
SN Ia+BAO/CMB& $66.02$ & $0.28$ &$-0.03$ & $0.05$ & $564.78$\\ 
\hline
\end{tabular}
\label{tab:fntable2} 
\end{center} 
\end{table}
\begin{table}[ht]
\caption{\em Best fit values of the model parameters $\alpha$ and $\beta$ (within $1\sigma$ confidence level) for the analysis of SN Ia+BAO+CMB (Planck) dataset by considering different values of $\Omega_{m0}$.}
\begin{center}
\begin{tabular}{|c|c|c|c|c|}
\hline
Data & $\Omega_{m0}$ & $\alpha$ & $\beta$ & $\chi^{2}_{m}$ \\ \hline
SN Ia+BAO+CMB (Planck) & $0.3$ &$-0.04$ & $0.03$ & $564.49$\\
& $0.315$ &$-0.05$ & $0.03$ & $564.32$\\ 
\hline
\end{tabular}
\label{tab:fntable3} 
\end{center} 
\end{table}
\begin{table}[ht]
\caption{\em Best fit values of $H_{0}$, $\Omega_{m0}$, $\alpha$ and $\beta$ (within $1\sigma$ confidence level) for the analysis of SN Ia+BAO+CMB (Planck) dataset.}
\begin{center}
\begin{tabular}{|c|c|c|c|c|c|}
\hline
Data & $H_{0}$ &$\Omega_{m0}$ & $\alpha$ & $\beta$ & $\chi^{2}_{m}$ \\ \hline
SN Ia+BAO+CMB (Planck)& $66.83$ & $0.294$ &$-0.05$ & $0.04$ & $565.79$\\ 
\hline
\end{tabular}
\label{tab:fntable4} 
\end{center} 
\end{table}
\par In the upper panel of figure \ref{figqz}, the evolution of the deceleration parameter $q(z)$ is shown within $1\sigma$ and $2\sigma$ confidence regions around the best fit curve for the combined dataset. It is clear from figure \ref{figqz} that $q(z)$ shows a smooth transition from a decelerated ($q > 0$, at high $z$) to an accelerated ($q < 0$, at low $z$) phase of expansion of the universe at the transition redshift $z_{t}=0.75$ for the best-fit model (as shown by central dark line). It deserves mention here that the value of $z_{t}$ obtained in the present work is very close to the value obtained for various dark energy models by Magana et al. \cite{jmang}. They have found that the universe has a transition from a decelerated phase to an accelerated phase at  $z_{t}\sim 0.75$, $z_{t}\sim 0.7$, $z_{t}\sim 1$ and $z_{t}\sim 0.7$ for the  Polynomial, BA, FSLL I and FSLL II parametrizations of $\omega_{\phi}(z)$ respectively (see \cite{jmang} and references there in). Also, the present value of $q$ (say, $q_{0}$) obtained in this work for the best-fit model is $-0.58$. Hence, the values of $z_{t}$ and $q_{0}$ obtained in the present work are very close to the value obtained for the standard $\Lambda$CDM model ($z_{t}\approx 0.74$ and $q_{0}\approx-0.59$), as indicated by the red dashed line in the upper panel of figure \ref{figqz}. Recently Ishida et al. \cite{baocmb2} used a kink-like expression for $q(z)$ to study the expansion history of the universe. They have obtained $z_{t}=0.84^{+0.13}_{-0.17}$ and $z_{t}=0.88^{+0.12}_{-0.10}$  (at $2\sigma$ confidence level) for $SDSS+2dfGRS ~BAO+Gold182$ and $SDSS+2dfGRS~BAO+SNLS$ datasets respectively. So, our analysis ($z_{t}=0.75\pm 0.02$, at $2\sigma$ level) provides better constraint on $z_{t}$ as compared to the results of Ishida et al. \cite{baocmb2}. Next, we have shown the reconstructed evolution history of the effective EoS parameter $\omega_{eff}(z)$ in the lower panel of figure \ref{figqz} for this model using SN Ia+BAO/CMB dataset. The lower panel of figure \ref{figqz} reveals that $\omega_{eff}(z)$ was very close to zero at high $z$ and attains negative value ($-1<\omega_{eff}<-\frac{1}{3}$, within $2\sigma$ limit) at low $z$, and thus does not suffer from the problem of `future singularity'. These results are also in good agreement with the observational data. We have also reconstructed the EoS parameter $\omega_{\phi}(z)$ for the scalar field in the inset diagram of the lower panel of figure \ref{figqz}. For the best-fit model, the present value of $\omega_{\phi}(z)$ comes out to be $-0.99^{+0.04}_{-0.03}$ (with $1\sigma$ errors) and $-0.99^{+0.08}_{-0.07}$ (with $2\sigma$ errors), which is definitely within the constraint range \cite{wv,dav}. Moreover, our results are also in good agreement with other previous works \cite{aam,jmang,kshi,aam2}, where the authors have considered different parameterizations of $\omega_{\phi}(z)$ and obtained $\omega_{\phi}(z=0)\approx -1$ at $1\sigma$ confidence level for the analysis of various observational datasets.\\  
\begin{figure}[ht]
\begin{center}
\includegraphics[width=0.38\textwidth,height=0.21\textheight]{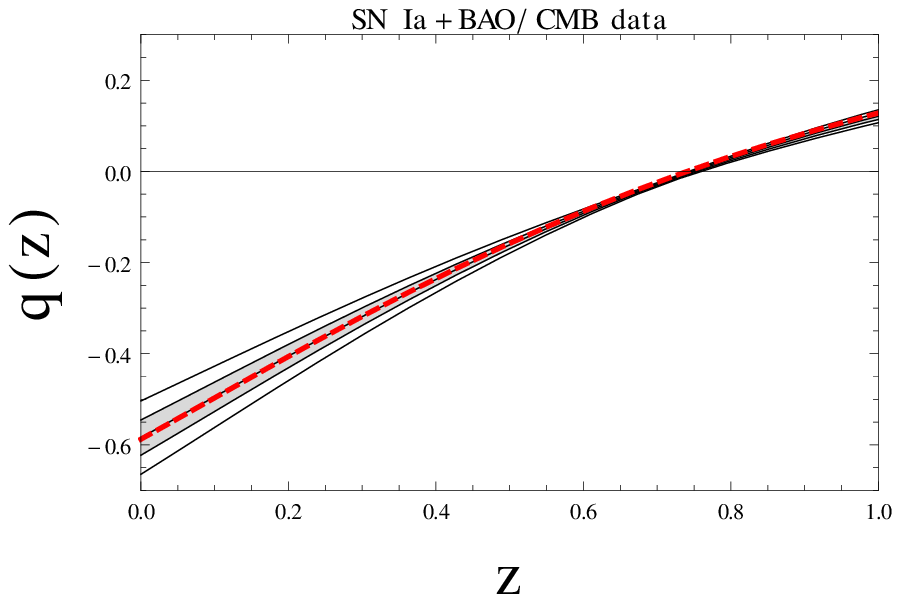}\\
\vspace{7mm}
\includegraphics[width=0.38\textwidth,height=0.21\textheight]{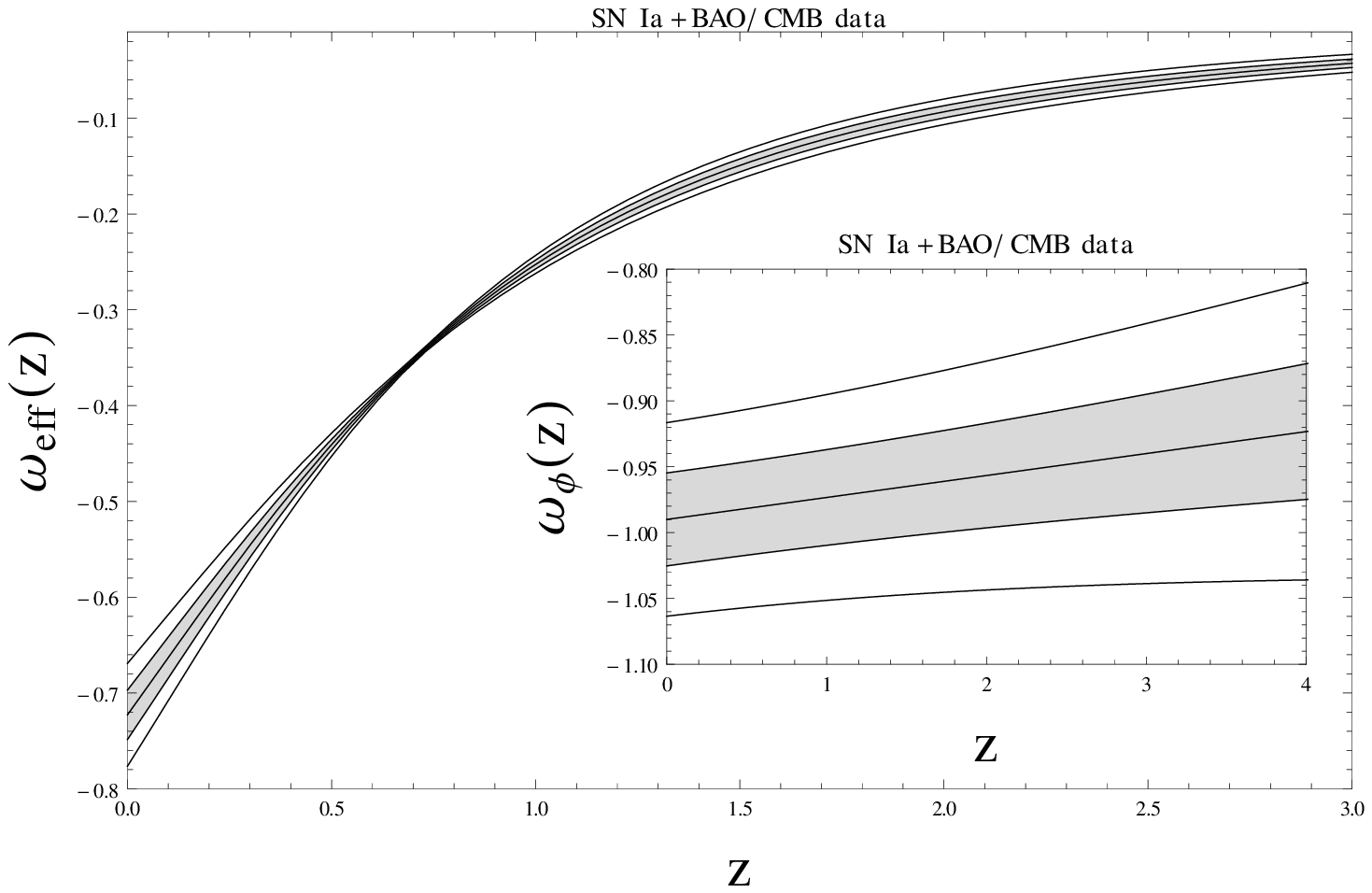}
\caption{\em a) Upper panel: The evolution of $q(z)$ as a function of $z$. The dashed line represents $\Lambda$CDM model with $\Omega_{\Lambda 0}=0.725$ and $\Omega_{m0}=0.275$. b) Lower panel: The evolution of $\omega_{eff}(z)$ as a function of $z$. Both the plots are for the best-fit values of the pair ($\alpha$, $\beta$) arising from the analysis of SN Ia+BAO/CMB dataset and $\Omega_{m0} = 0.27$ (see table \ref{tab:fntable1}). The $1\sigma$ and $2\sigma$ confidence regions have been shown and the central dark line represents the best fit curve.}
\label{figqz}
\end{center}
\end{figure}
\par The upper panel of figure \ref{figdmpot1} shows the evolution of the potential $V(z)$ as a function of $z$. The best fit of the potential, as indicated by the central line, remains almost constant in the range $0<z<3$. For the sake of completeness, using the parametric relations [$\phi(z)$, $V(z)$] given by equations (\ref{pz1}) and (\ref{vz1}), we have also obtained the form of the dark energy potential $V(\phi)$ by a numerical method for some given values of the model parameters. The evolution of $V(\phi)$ is shown in the lower panel of figure \ref{figdmpot1} and it has been found that $V(\phi)$ sharply increases with $\phi$ for the choice of $X(z)$ given by equation (\ref{eans1}). For this plot, we have considered $\alpha=-0.02$, $\beta=0.05$, $\Omega_{m0}=0.27$ and $\phi_{0}=0.1$. In this case, the potential $V(\phi)$ can be explicitly expressed in terms of $\phi$ as
\be\begin{split} &
\frac{V(\phi)}{3H^{2}_{0}}\approx 2579.13 \phi^{5}-2554.97\phi^{4}\\& 
~~~~~~~~~~~~~~~~~~~~~~+998.06\phi^{3} -189.63\phi^{2}+17.57\phi +0.1
\end{split}
\ee
which comes to be a polynomial in $\phi$.\\
\begin{figure}[ht]
\begin{center}
\includegraphics[width=0.38\textwidth,height=0.21\textheight]{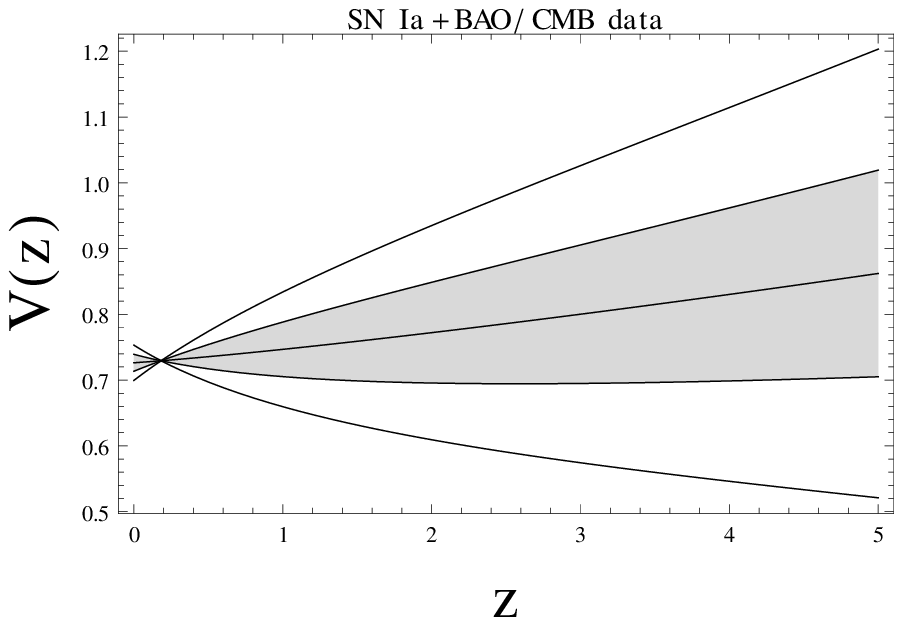}\\
\vspace{7mm}
\includegraphics[width=0.38\textwidth,height=0.21\textheight]{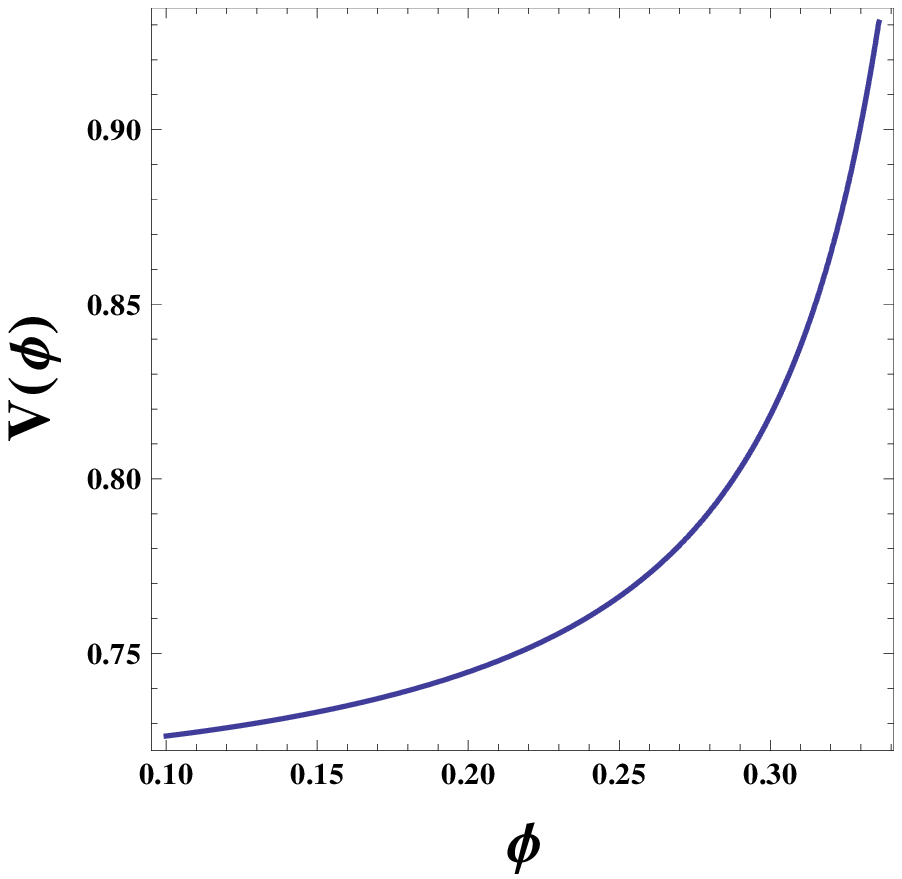}
\caption{\em a) Upper panel: The plot of the scalar potential $V(z)$ (in units of $3H^{2}_{0}$) as a function of $z$ with $1\sigma$ and $2\sigma$ confidence regions. The central dark line is the best fit curve. b) Lower panel: The evolution of the reconstructed potential (in units of $3H^{2}_{0}$) with the scalar field $\phi$ by considering $\phi_{0}=0.1$. Both the plots are for the best-fit values of ($\alpha ,\beta$) for the SNIa+BAO/CMB dataset and $\Omega_{m0}=0.27$. }
\label{figdmpot1}
\end{center}
\end{figure}
\par The upper panel of figure \ref{figxz} shows the evolution of $X(z)$ as a function of $z$ at the $1\sigma$ and $2\sigma$ confidence levels. It can be seen from figure \ref{figxz} that $X(z)$ behaves like cosmological constant (i.e., $X(z)=1$) at the present epoch, but deviation from this is clearly visible at high redshift. The variation of energy densities $\rho_{m}$ and $\rho_{\phi}$ with the redshift $z$ are also shown in the lower panel of figure \ref{figxz}, which shows that $\rho_{\phi}$ dominates over $\rho_{m}$ at the present epoch. This result is in accordance with observational predictions. 
\begin{figure}[ht]
\begin{center}
\includegraphics[width=0.38\textwidth,height=0.21\textheight]{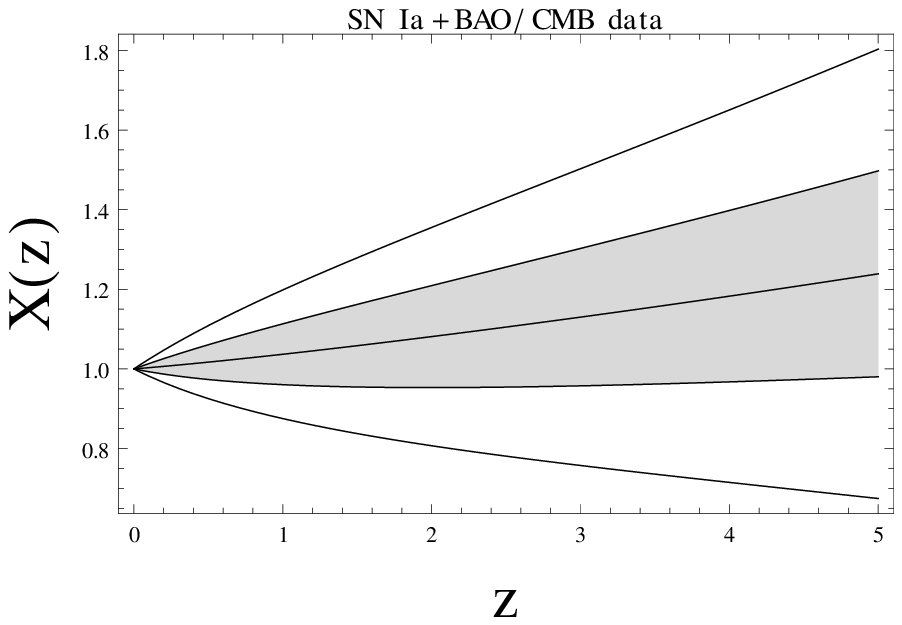}\\
\vspace{7mm}
\includegraphics[width=0.38\textwidth,height=0.21\textheight]{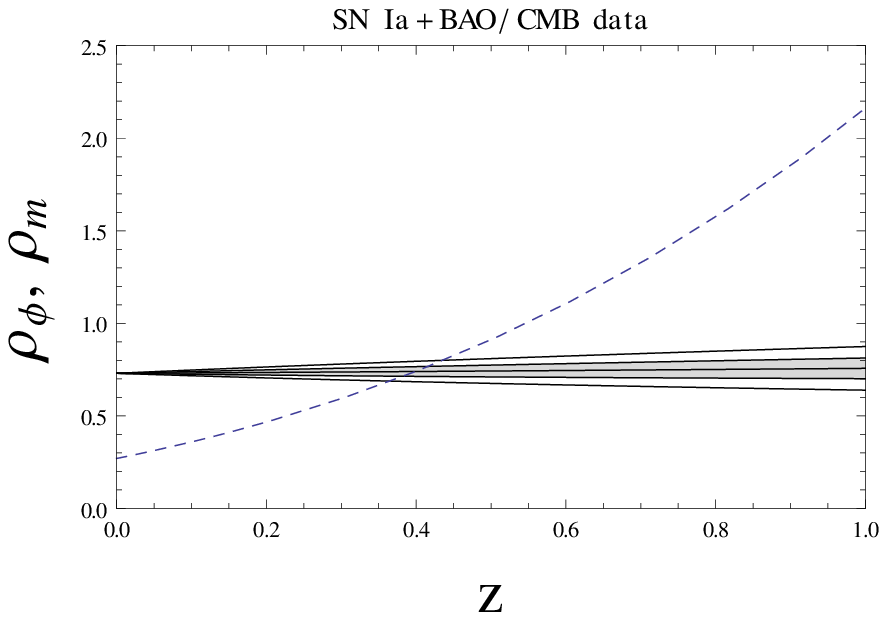}
\caption{\em Upper panel: The evolution of $X(z)$ with $1\sigma$ and $2\sigma$ confidence regions. Lower panel: Plot of $\rho_{m}(z)$ (dashed curve) and $\rho_{\phi}(z)$ (with $1\sigma$ and $2\sigma$ confidence regions) as a function of $z$ for this toy model with $\Omega_{m0}= 0.27$ (in units of $3H^{2}_{0}$). In each panel, the central dark line is the best fit curve.}
\label{figxz}
\end{center}
\end{figure}
\section{Conclusion}\label{conclusion}
In this paper, we have focused on a quintessence model in which the scalar field is considered as a candidate of dark energy. It has been shown that for a spatially flat FRW universe, we can construct a presently accelerating model of the universe with the history of a deceleration in the past by considering a specific choice of the dimensionless dark energy function $X(z)$. The motivation behind this particular choice of $X(z)$ has been discussed in details in section \ref{sec2} and for this specific ansatz, we have solved the field equations and have obtained the expressions for different cosmological parameters, such as $H(z)$, $q(z)$ and $\omega_{eff}(z)$. As mentioned earlier that the model parameters ($\alpha$ and $\beta$) are a good indicator of deviation of the present model from cosmological constant as for $\alpha=0$ and $\beta=0$ the model mimics the $\Lambda$CDM model. We have also constrained the model parameters using the SN Ia+BAO/CMB(WMAP7) and SN Ia+BAO+CMB(Planck) datasets to study the different properties of this model extensively. It is evident from table \ref{tab:fntable1} that the best-fit values of $\alpha$ and $\beta$ are very close to zero. So, our analysis indicates that the reconstructed $\omega_{\phi}(z)$ is very close to the $\Lambda$CDM value at the present epoch. In summary, using SN Ia+BAO/CMB dataset jointly, we have then reconstructed various parameters (e.g., $q(z)$, $\omega_{eff}(z)$ and $\omega_{\phi}(z)$) as well as the quintessence potential $V(\phi)$ directly from the chosen $X(z)$, which describes the properties of the dark energy. The resulting cosmological scenarios are found to be very interesting. It has been found that the evolution of $q(z)$ in this model shows a smooth transition from a decelerated to an accelerated phase of expansion of the universe at late times. As discussed in section \ref{result}, it has been found that our reconstructed results of $q(z)$ and $\omega_{\phi}(z)$ are in good agreement with the previous works \cite{aam,jmang,kshi,aam2}. For completeness of the work, we have also derived the form of the effective scalar field potential $V(\phi)$, in terms of $\phi$, for this model and the resulting potential is found to be a polynomial in $\phi$. \\
\par From the present investigation, it can be concluded that the SN Ia+BAO/CMB dataset although supports the concordance $\Lambda$CDM model at the $1\sigma$ confidence level, but it favors the scalar field dark energy model as well. In other words, it is well worth emphasizing that the observational datasets are not yet good enough to strongly distinguish present dark energy model from the $\Lambda$CDM model at present. With the progress of the observational techniques as well as the data analysis methods in the future, we hope that the parameters in $X(z)$ can be constrained more precisely, which will improve our understanding about the nature of dark energy. The present analysis is one preliminary step towards that direction. In future, we plan to test this parametric form of $X(z)$ in scalar-tensor theories of gravity. 
\section{Acknowledgements} A.A.M. acknowledges UGC, Govt. of India for financial support through Maulana Azad National Fellowship. This work was partially supported by the JSPS Grant-in-Aid for Young Scientists (B) \# 25800136 and the research-funds presented by Fukushima University (K.B.). S.D. wishes to thank IUCAA, Pune for associateship program. Authors are also thankful to the anonymous referee whose useful suggestions have improved the quality of the paper.

\end{document}